\documentclass{llncs}

\usepackage{bm,cite,float,graphicx,multirow,subfigure}
\usepackage[colorlinks,bookmarksopen,bookmarksnumbered,citecolor=red,urlcolor=red]{hyperref}
\def\etal{\emph{et al.}~}

\begin{document}

\title{A Convolutional Neural Network Approach for Post-Processing in HEVC Intra Coding}

\author{Yuanying Dai, Dong Liu, and Feng Wu}
\institute{CAS Key Laboratory of Technology in Geo-Spatial Information Processing and Application System, University of Science and Technology of China, Hefei, China, \email{daiyy@mail.ustc.edu.cn, dongeliu@ustc.edu.cn, fengwu@ustc.edu.cn}}

\maketitle

\begin{abstract}
Lossy image and video compression algorithms yield visually annoying artifacts including blocking, blurring, and ringing, especially at low bit-rates. To reduce these artifacts, post-processing techniques have been extensively studied. Recently, inspired by the great success of convolutional neural network (CNN) in computer vision, some researches were performed on adopting CNN in post-processing, mostly for JPEG compressed images. In this paper, we present a CNN-based post-processing algorithm for High Efficiency Video Coding (HEVC), the state-of-the-art video coding standard. We redesign a Variable-filter-size Residue-learning CNN (VRCNN) to improve the performance and to accelerate network training. Experimental results show that using our VRCNN as post-processing leads to on average 4.6\% bit-rate reduction compared to HEVC baseline. The VRCNN outperforms previously studied networks in achieving higher bit-rate reduction, lower memory cost, and multiplied computational speedup.
\end{abstract}


\section{Introduction}
\label{sec:intro}

Lossy image and video compression algorithms, such as JPEG \cite{wallace1992jpeg} and High Efficiency Video Coding (HEVC) \cite{sullivan2012overview}, by nature cause distortion and yield artifacts especially at low bit-rates. For example, due to block-based coding, there are visible discontinuities at block boundaries in compressed images, which are known as blocking artifacts; due to loss of high-frequency components, compressed images often become blurred than the original. Other artifacts include ringing, color bias, and so on. These compression artifacts may severely decrease the perceptual quality of reconstructed image or video, and thus how to reduce or remove these artifacts is an important problem and has been extensively studied in the literature.

In HEVC, the state-of-the-art video coding standard, there are two post-processing techniques for artifact reduction, namely deblocking \cite{norkin2012hevc} and sample adaptive offset (SAO) \cite{fu2012sample}. The differences between deblocking and SAO are twofold. First, deblocking is specifically designed to reduce blocking artifacts, but SAO is designed for general compression artifacts. Second, deblocking does not require any additional bit, but SAO requires to transmit some additional bits for signaling the offset values. Both techniques contribute to the improvement of the visual quality of reconstructed video, and also help to improve the objective quality and equivalently achieve bit-rate saving.

Recently, convolutional neural network (CNN) achieved great success in high-level computer vision tasks such as image classification \cite{krizhevsky2012imagenet} and object detection \cite{girshick2014rich}. Inspired by the success, it was also proposed to utilize CNN for low-level computer vision tasks such as super-resolution \cite{dong2014learning,kim2016accurate} and edge detection \cite{xie2015holistically}.

More recently, Dong \etal proposed an artifact reduction CNN (AR-CNN) \cite{dong2015compression} approach for reducing artifacts in JPEG compressed images. The AR-CNN is built upon their previously designed super-resolution CNN (SRCNN) \cite{dong2014learning}, and reported to achieve more than 1 dB improvement over JPEG images. Wang \etal \cite{wang2016d3} investigated another network structure for JPEG artifact reduction. Furthermore, Park and Kim \cite{park2016cnn} proposed to utilize the SRCNN network to replace the deblocking or SAO in HEVC, and reported achieving bit-rate reduction. However, the results in \cite{park2016cnn} were achieved by training a network with several frames of a video sequence and then testing the network with the same sequence, which cannot reveal the generalizability of the trained network.

In this paper, we present a redesigned CNN for artifact reduction in HEVC intra coding. We propose to integrate variable filter size into the designed CNN to improve its performance. We also utilize the recently proposed residue learning technique \cite{he2016deep} to accelerate the training of CNN. Moreover, we trained the network with a collection of natural images and tested the network with the standard video sequences, so as to demonstrate the generalizability of the network. Our proposed Variable-filter-size Residue-learning CNN (VRCNN) can be adopted as post-processing to replace deblocking and SAO, as it reduces general compression artifacts and requires no additional bit. Experimental results show that VRCNN achieves on average 4.6\% bit-rate reduction compared to deblocking and SAO in HEVC baseline. The VRCNN also outperforms previously studied networks in achieving higher bit-rate reduction, lower memory cost, and multiplied computational speedup.

The remainder of this paper is organized as follows. Section \ref{sec:formation} presents the details of the designed VRCNN. Section \ref{sec:Implementation} discusses the details of training and using VRCNN. Section \ref{sec:result} gives out the experimental results, followed by conclusions in Section \ref{sec:conclusion}.
\section{Our Designed CNN}
\label{sec:formation}
Currently, there are several existing networks for artifact reduction: AR-CNN \cite{dong2015compression}, D$^3$ \cite{wang2016d3}, and SRCNN \cite{park2016cnn}. Note that AR-CNN is built upon SRCNN, and the SRCNN was originally designed for super-resolution \cite{dong2014learning}. The D$^3$ network was specifically designed for JPEG as it utilized the JPEG built-in 8$\times$8 discrete cosine transform (DCT), and thus not suitable for HEVC which adopts variable block size transform. In the following, we first discuss on AR-CNN, and then presents our redesigned VRCNN.
\subsection{AR-CNN}
AR-CNN is a 4-layer fully convolutional neural network. It has no pooling or full-connection layer, so the output can be of the same size as the input given proper boundary condition (the boundary condition of convolutions will be discussed later). Denote the input by $\bm{Y}$, the output of layer $i\in\{1,2,3,4\}$ by $F_i(\bm{Y})$, and the final output by $F(\bm{Y})=F_4(\bm{Y})$, then the network can be represented as:
\begin{eqnarray}
  F_1(\bm{Y}) &=& g(W_1*\bm{Y}+B_1) \\
  F_i(\bm{Y}) &=& g(W_i*F_{i-1}(\bm{Y})+B_i), i\in\{2,3\} \\
  F(\bm{Y}) &=& W_4*F_3(\bm{Y})+B_4
\end{eqnarray}
where $W_i$ and $B_i$ are the weights and biases parameters of layer $i$, $*$ stands for convolution, and $g()$ is a non-linear mapping function. In recent CNNs, the rectified linear unit (ReLU) \cite{nair2010rectified} is often adopted as the non-linear mapping, i.e. $g(x)=\max(0,x)$.

\begin{table}
\begin{center}
\caption{The configuration of AR-CNN \cite{dong2015compression}} \label{tab:para-arcnn}
\begin{tabular}{|l|c|c|c|c|}
 \hline
Layer&Layer 1&Layer 2&Layer 3&Layer 4\\
 \hline
Filter size & 9$\times$9 & 7$\times$7&1$\times$1&5$\times$5\\
 \hline
\# filters &64 & 32&16&1\\
 \hline
\# parameters & 5184 & 100352&512&400\\
 \hline
 Total parameters & \multicolumn{4}{c|}{106448}\\
\hline
\end{tabular}
\end{center}
\end{table}
The four layers in AR-CNN are claimed to perform four steps of artifact reduction: feature extraction, feature enhancement, mapping, and reconstruction (as discussed in \cite{dong2015compression}). Accordingly, the configuration of AR-CNN is summarized in Table \ref{tab:para-arcnn}. Note that the amount of (convolutional) parameters in each layer is calculated as (number of filters in the last layer)$\times$(number of filters in this layer)$\times$(filter size).
\subsection{VRCNN}
\begin{figure}
   \centering
    {
    \includegraphics[width=\textwidth]{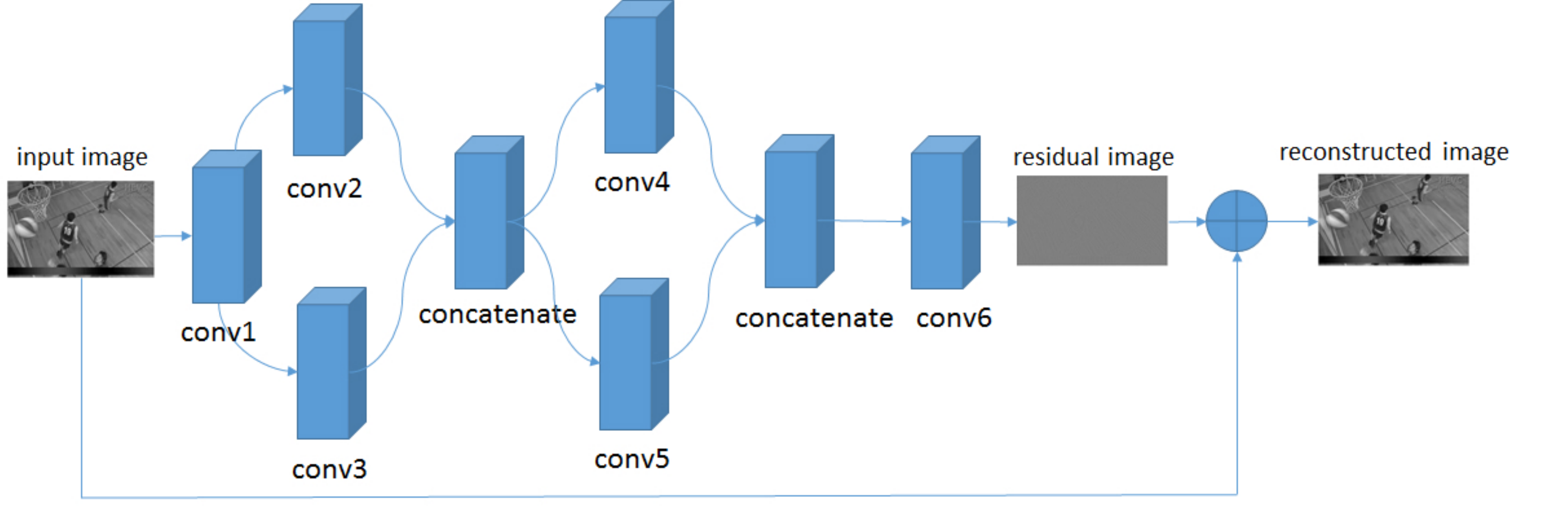}
    }
	\caption{The structure of VRCNN, a 4-layer fully convolutional neural network.}
	\label{fig:network}
\end{figure}
\begin{table}
\begin{center}
\caption{The configuration of VRCNN} \label{tab:para}
\begin{tabular}{|l|c|c|c|c|c|c|}
 \hline
Layer&Layer 1&\multicolumn{2}{c|}{Layer 2}&\multicolumn{2}{c|}{Layer 3}&Layer 4\\
\hline
Conv. module&conv1&conv2&conv3&conv4&conv5&conv6\\
 \hline
Filter size & 5$\times$5 & 5$\times$5&3$\times$3&3$\times$3&1$\times$1&3$\times$3\\
 \hline
\# filters &64 & 16&32&16&32&1\\
 \hline
\# parameters & 1600 & 25600&18432&6912&1536&432\\
 \hline
 Total parameters & \multicolumn{6}{c|}{54512}\\
\hline
\end{tabular}
\end{center}
\end{table}
Since AR-CNN is designed for JPEG, but our aim is to perform artifact reduction for HEVC, we redesign the CNN structure and name it VRCNN. The structure of VRCNN is shown in Fig. \ref{fig:network} and its configuration is given in Table \ref{tab:para}. It is also a 4-layer fully convolutional neural network, like AR-CNN. We now discuss the differences between VRCNN and AR-CNN.

The rationale cause of compression artifacts in JPEG and HEVC is the quantization of transformed coefficients. The transform is block wise, thus the quantization error of one coefficient affects only the pixels in the same block. As JPEG adopts fixed 8$\times$8 DCT, but HEVC adopts variable block size transform \footnote{HEVC adopts 4$\times$4, 8$\times$8, 16$\times$16, up to 32$\times$32 DCT, and allows the choice of discrete sine transform (DST) at 4$\times$4.}, which shall be taken into account to reduce the quantization error. Therefore, we propose to adopt variable filter size in the second layer, because this layer is designed to make the ``noisy'' features ``cleaner'' \cite{dong2015compression}. Specifically, we replace the second layer of AR-CNN (fixed 7$\times$7 filters) with the combination of 5$\times$5 and 3$\times$3 filters. The outputs of different-sized filters are concatenated to be fed into the next layer. Similarly, we also adopt variable filter size in the third layer that performs ``restoration'' of features \cite{dong2015compression}. The fixed 1$\times$1 filters in AR-CNN are replaced by combination of 3$\times$3 and 1$\times$1 filters. Note that the first and the last layers of VRCNN do not use variable filter size, because these two layers perform feature extraction and final reconstruction, respectively \cite{dong2015compression}, which are not affected by variable block size transform of HEVC.

The variable filter size technique, i.e. combination of filters of different sizes in one layer of CNN, has been proposed earlier in CNNs for image classification, e.g. the well-known GoogleNet \cite{szegedy2015going}, where different-sized filters are to provide multi-scale information of the input image. In our VRCNN, variable filter size is proposed to suit for HEVC variable block size transform, and thus used only in selected layers. To the best of our knowledge, VRCNN is the first network that uses variable filter size for artifact reduction.

In addition, we propose to integrate the recently developed residue learning technique \cite{he2016deep} into VRCNN. That is, the output of the last layer is added back to the input, and the final output is:
\begin{equation}
  F(\bm{Y}) = W_4*F_3(\bm{Y})+B_4+\bm{Y}
\end{equation}
In other words, the CNN is designed to learn the residue between output and input rather than directly learning the output. In the case of artifact reduction, the input (before filtering) and the output (after filtering) shall be similar to the other to a large extent, therefore, learning the difference between them can be easier and more robust. Our empirical study indeed confirms that residue learning converges much faster. Note that residue learning is also a common strategy in super-resolution with or without CNN \cite{kim2016accurate,sun2003image}.

Last but not the least, to integrate CNN into the in-loop post-processing of HEVC, it is very important to control the network complexity. For that purpose, our designed VRCNN is greatly simplified than AR-CNN. Comparing Table \ref{tab:para-arcnn} and Table \ref{tab:para}, VRCNN uses more filters, but at smaller sizes. As a result, the amount of parameters is greatly reduced in VRCNN. We notice that recent work on super-resolution also uses smaller filters but much more (20) layers \cite{kim2016accurate}, while VRCNN has 4 layers like AR-CNN.
\section{Training and Using VRCNN}
\label{sec:Implementation}
We propose to adopt VRCNN for post-processing in HEVC to replace the original deblocking and SAO. In order to make a fair comparison with the original deblocking and SAO, we train the VRCNN on a collection of natural images, and test it on the HEVC standard test sequences. The training and testing images have no overlap so as to demonstrate the generalizability of the trained network.
\subsection{Training}
An original image $\bm{X}_n$, where $n\in\{1,\dots,N\}$ indexes each image, is compressed with HEVC intra coding, while turning off deblocking and SAO, and the compressed image is regarded as the input to VRCNN, i.e. $\bm{Y}_n$. The objective of training is to minimize the following loss function:
\begin{equation}\label{loss}
  L(\Theta)=\frac{1}{N}\sum_{n=1}^{N}{{\left\|F(\bm{Y}_n|\Theta)-\bm{X}_n\right\|}^2}
\end{equation}
where $\Theta$ is the whole parameter set of VRCNN, including $W_i,B_i,i\in\{1,2,3,4\}$. This loss is minimized using stochastic gradient descent with the standard back-propagation.

In order to accelerate the training, we also adopt the adjustable gradient clipping technique proposed in \cite{kim2016accurate}. That is, the learning rate $\alpha$ is set large, but the actual gradient update is restricted to be in the range of $[-\tau/\alpha,\tau/\alpha]$ where $\tau$ is a constant (set to 0.01 in our experiments). The key idea beneath this technique is to clip the gradient when $\alpha$ is large, so as to avoid exploding. As training goes on, the learning rate $\alpha$ becomes smaller and then the range is too large to be actually used.
\subsection{Using VRCNN}
We integrate a trained VRCNN into HEVC intra coding. The deblocking and SAO are turned off, and the compressed intra frame is directly fed into the trained VRCNN, producing the final reconstructed frame. Unlike SAO, the VRCNN needs no additional bit, but still can reduce general compression artifacts as demonstrated by experimental results. Therefore, in all-intra coding setting, VRCNN can be made in-loop or out-of-loop. One remaining issue is the boundary condition for convolutions. In this work, we follow the practice in \cite{kim2016accurate}, i.e. padding zeros before each convolutional module so that the output is of the same size as the input. Zero-padding seems quite simple but works well in experiments.

\section{Experimental Results}
\label{sec:result}
\subsection{Implementation}
\label{ssec:dataset}
We use the software Caffe \cite{jia2014caffe} for training VRCNN as well as comparative networks on a NVIDIA Tesla K40C graphical processing unit (GPU). A collection of 400 natural images, i.e. the same set as that in \cite{dong2015compression}, are used for training. Each original image is compressed by HEVC intra coding (deblocking and SAO turned off) at four different quantization parameters (QPs): 22, 27, 32, and 37. For each QP, a separate network is trained out. Only the luminance channel (i.e. Y out of YUV) is considered for training. Due to the limited memory of the GPU, we do not use the entire image as a sample. Instead, the original image $\bm{X}_n$ and compressed image $\bm{Y}_n$ are both divided into 35$\times$35 sub-images without overlap. The corresponding pair of sub-images is regarded as a sample, so we have in total 46,784 training samples. Note that different from \cite{dong2015compression}, we use zero padding before each convolution so that the output is of the same size as the input, and therefore the loss is computed over the entire sub-image.
\begin{table}
\begin{center}
\caption{The BD-rate results of our VRCNN compared to HEVC baseline} \label{Tab:VRCNN_anchor}
\begin{tabular}{|l|l|c|c|c|}
 \hline
  \multirow{2}{*}{Class}& \multirow{2}{*}{Sequence}&\multicolumn{3}{c|}{BD-rate}\\
 \cline{3-5}
 &&Y (\%) & U (\%) &V (\%)\\
 \hline
  \multirow{4}{*}{Class A}&Traffic & -5.6&-3.5 &-4.1\\
 \cline{2-5}
 &PeopleOnStreet & -5.4 &-5.9 &-5.7\\
  \cline{2-5}
 &Nebuta & -0.9 &-4.9 &-4.1\\
  \cline{2-5}
 &SteamLocomotive &-1.9 &-0.5 &-0.3\\
 \hline
 \multirow{5}{*}{Class B} & Kimono & -2.5&-1.5 &-1.4\\
  \cline{2-5}
&ParkScene & -4.4 &-3.3 &-2.5\\
  \cline{2-5}
 &Cactus & -4.6 &-3.9 &-6.3\\
 \cline{2-5}
&BasketballDrive & -2.5 &-3.7 &-5.3\\
 \cline{2-5}
&BQTerrace & -2.6 &-3.3 &-3.0\\
 \hline
\multirow{4}{*}{Class C}& BasketballDrill &-6.9  &-5.8 &-6.8\\
  \cline{2-5}
 &BQMall & -5.1 &-5.3 &-5.3\\
  \cline{2-5}
&PartyScene &-3.6  &-4.4 &-4.4\\
\cline{2-5}
&RaceHorses &-4.2  &-6.7 &-11.0\\
 \hline
 \multirow{5}{*}{Class D} & BasketballPass & -5.3&-4.4 &-6.5\\
 \cline{2-5}
&BQSquare &  -3.8&-4.2 &-6.4\\
 \cline{2-5}
 &BlowingBubbles & -4.9 &-8.4 &-7.9\\
 \cline{2-5}
&RaceHorses & -7.6 &-8.5 &-11.5\\
 \hline
\multirow{4}{*}{Class E}& FourPeople & -7.0 &-5.3 &-5.2\\
 \cline{2-5}
 &Johnny & -5.9 &-5.0 &-5.5\\
 \cline{2-5}
&KristenAndSara & -6.7 &-6.1 &-6.2\\
\hline
 \multirow{5}{*}{Class Summary}
 &Class A&-3.5&-3.7&-3.6\\
  \cline{2-5}
 &Class B&-3.3&-3.2&-3.7\\
  \cline{2-5}
 &Class C&-5.0&-5.5&-6.9\\
  \cline{2-5}
 &Class D&-5.4&-6.4&-8.1\\
  \cline{2-5}
 &Class E&-6.5&-5.5&-5.6\\
 \hline
\textbf{Overall} &\textbf{All}& \textbf{-4.6} & \textbf{-4.7}&\textbf{-5.5}\\
 \hline
\end{tabular}
\end{center}
\end{table}

During network training, the weights are initialized using the method in \cite{he2016deep}. Training samples are randomly shuffled and the mini-batch size is 64. The momentum parameter is set to 0.9, and weight decay is 0.0001. The base learning rate is set to decay exponentially from 0.1 to 0.0001, changing every 40 epochs. Thus, in total the training takes 160 epochs and uses around 1.5 hours on our GPU. The bias learning rate is set to 0.01, 0.01, and 0.1, for QP 27, 32, and 37, respectively. For QP 22, the network is not trained from scratch but rather fine-tuned from the network of QP 27. For this fine tuning, the base learning rate is 0.001, bias learning rate is 0.0001, and training finishes after 40 epochs.

For comparison, we also trained other two networks, AR-CNN \cite{dong2015compression} and VDSR \cite{kim2016accurate}, using the same training images. The AR-CNN is designed for JPEG artifact reduction, so we cannot reuse their trained network for HEVC, but we re-train the network from scratch using the source code provided by the authors. The training proceeds in 2,500,000 iterations and takes almost 5 days. The VDSR is proposed for super-resolution and claimed to outperform SRCNN (the basis of AR-CNN), so we also include it for comparison. We also re-train the network from scratch and manually tune the training hyper-parameters. The training takes about 6 hours to finish 38,480 iterations. It can be observed that the training of our VRCNN and VDSR is significantly faster than that of AR-CNN, because both VRCNN and VDSR adopt residue learning. And the training of our VRCNN is also faster than VDSR since our network is much more simple.

After training, we integrate the network into HEVC reference software HM \footnote{HM version 16.0, \url{https://hevc.hhi.fraunhofer.de/svn/svn_HEVCSoftware/tags/HM-16.0/}.} and test on the HEVC standard test sequences. Five classes, 20 sequences are used for test. Class F is not used as it is screen content. For each sequence, only the first frame is used for test. Four QPs are tested: 22, 27, 32, 37, and for each QP the corresponding network is used. For AR-CNN and VDSR, we also train a separate network for each QP. Note that the training is performed on the luminance channel (Y) but the trained network is also used for chrominance channels (U and V).

Compared to \cite{park2016cnn}, which uses the same sequences for training and test, our experiments can better reveal the generalizability of the trained network and make fair comparison with the original deblocking and SAO. Moreover, the results are not complete for all HEVC standard test sequences \cite{park2016cnn}. Therefore, we do not include its results in the following.
\subsection{Comparison with HEVC Baseline}
We first compare our VRCNN used as post-processing against the original deblocking and SAO. To evaluate the coding efficiency, we use the BD-rate measure \cite{bjontegaard2001calcuation} on luminance and chrominance channels independently. The results are summarized in Table \ref{Tab:VRCNN_anchor}. It can be observed that the VRCNN achieves significant bit-rate reduction on all the test sequences. For the luminance (Y), as high as 7.6\% BD-rate is achieved on the \texttt{RaceHorses} sequence, and on average 4.6\% BD-rate is achieved on all the sequences. For the chrominance (U and V), the BD-rate is more significant for several sequences, reaching as high as 11.5\% on the \texttt{RaceHorses} sequence. Note that the network is trained only on the luminance channel, this result shows that the network can be readily used for the chrominance channels, too.
\begin{figure}
  \centering
  \subfigure[Original]{\includegraphics[width=.45\textwidth]{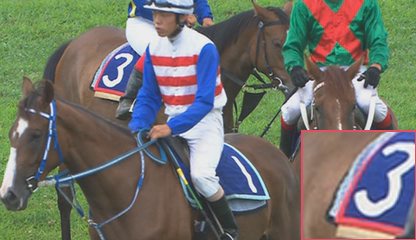}}
  \subfigure[Before post-processing. PSNR: 31.4460 dB]{\includegraphics[width=.45\textwidth]{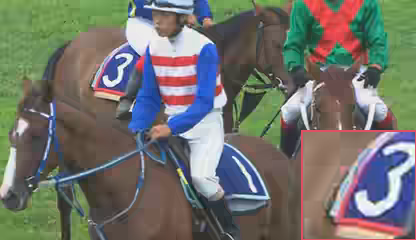}}
  \subfigure[HEVC baseline. PSNR: 31.6604 dB]{\includegraphics[width=.45\textwidth]{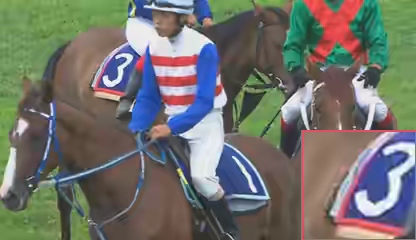}}
  \subfigure[AR-CNN. PSNR: 32.0764 dB]{\includegraphics[width=.45\textwidth]{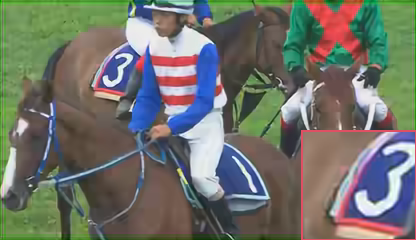}}
  \subfigure[VDSR. PSNR: 32.1050 dB]{\includegraphics[width=.45\textwidth]{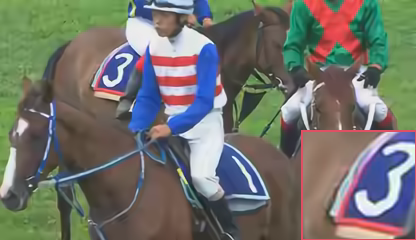}}
  \subfigure[VRCNN (ours). PSNR: 32.2413 dB]{\includegraphics[width=.45\textwidth]{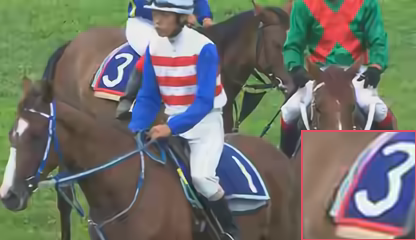}}
  \caption{The first frame of \texttt{RaceHorses}, compressed at QP 37, and post-processed by HEVC baseline as well as different CNNs.}\label{fig:subjective}
\end{figure}

We also compare the visual quality of reconstructed images as shown in Fig. \ref{fig:subjective}. A portion of each image is enlarged as inset in the bottom-right corner of each image. It can be observed that the image before post-processing \ref{fig:subjective} (b) contains obvious blocking and ringing artifacts. The processed image by HEVC baseline \ref{fig:subjective} (c) greatly reduces blocking, but ringing is still visible. The processed image by VRCNN \ref{fig:subjective} (f) suppresses all kinds of artifacts and produces better visual quality than \ref{fig:subjective} (c).
\subsection{Comparison with Other Networks}
\begin{table}
\begin{center}
\caption{The BD-rate results of AR-CNN and VDSR compared to HEVC baseline} \label{Tab:ARCNN_VDSR_anchor}
\begin{tabular}{|l|l|c|c|c|}
 \hline
  \multicolumn{2}{|c|}{\multirow{2}{*}{Network}} & \multicolumn{3}{c|}{\multirow{1}{*}{BD-rate}} \\
  \cline{3-5}
 \multicolumn{2}{|c|}{}&Y (\%) & U (\%) &V (\%)\\
 \hline
 \multirow{5}{*}{AR-CNN}&Class A& 0.9 &2.1 &2.1 \\
 \cline{2-5}
 &Class B&1.0&3.3&4.5\\
 \cline{2-5}
 &Class C&-0.6&2.6&4.0\\
 \cline{2-5}
 &Class D&-0.8&1.9&2.0\\
 \cline{2-5}
 &Class E&0.4&5.5&6.1\\
 \cline{2-5}
 &\textbf{Overall}&\textbf{0.2}&\textbf{3.0}&\textbf{3.7}\\
 \hline
 \multirow{5}{*}{VDSR}&Class A& -2.8 &-3.2 &-3.1 \\
 \cline{2-5}
 &Class B&-2.7&-2.7&-3.3\\
 \cline{2-5}
 &Class C&-4.1&-4.8&-5.7\\
 \cline{2-5}
 &Class D&-4.4&-5.6&-7.3\\
 \cline{2-5}
 &Class E&-5.7&-5.7&-6.1\\
 \cline{2-5}
 &\textbf{Overall}&\textbf{-3.8}&\textbf{-4.3}&\textbf{-4.9}\\
 \hline
\end{tabular}
\end{center}
\end{table}
We also compare our VRCNN with AR-CNN and VDSR to demonstrate the advantage of our redesigned network structure. First, the coding efficiency of each network is evaluated using the BD-rate measure. The results are summarized in Table \ref{Tab:ARCNN_VDSR_anchor}. It can be observed that AR-CNN performs slightly worse than the original deblocking and SAO in HEVC baseline, but VDSR also demonstrates significant gain. Note that VDSR is proposed for super-resolution and claimed to outperform SRCNN, while AR-CNN is built upon SRCNN, this result is reasonable because VDSR is much deeper (20 layers) than AR-CNN (4 layers). However, our proposed VRCNN, being also 4-layer, outperforms AR-CNN significantly, and also outperforms VDSR slightly, in terms of BD-rate. Since our VRCNN features variable filter size and residue learning compared to AR-CNN, this result demonstrates that carefully designed shallow network may still be competitive with deep network for artifact reduction.

The reconstructed images using AR-CNN and VDSR are also shown in Fig. \ref{fig:subjective} for comparison. The image obtained by AR-CNN contains slight blocking artifacts, but the image obtained by VDSR and our VRCNN have eliminated most compression artifacts. The visual quality comparison is consistent with the objective BD-rate measure.

We also compare the computational complexity of different networks. This comparison was performed on a personal computer with Intel core i7-4790K central processing unit (CPU) at 4GHz and NVIDIA GeForce GTX 750Ti GPU with 2GB memory. Due to the limited memory of GPU, we cannot process large images using Caffe's GPU mode on this computer. Thus we used the sequence \emph{Suzie} at resolution 176$\times$144, the first 10 frames are used for test under all-intra setting, and the decoding time results are summarized in Table \ref{Tab:time}. Note that Caffe can work in CPU or GPU mode, both modes are tested. The reported decoding time includes both CPU computation and GPU computation if have. Since most computations of decoding are performed by CPU, post-processing, if using Caffe's GPU mode, is the last step, thus the transmission time between CPU and GPU is not negligible. Overall, it can be observed that our VRCNN is more than 2$\times$ faster than VDSR, since the latter is much deeper. Moreover, though VRCNN and AR-CNN are both 4-layer, VRCNN is slightly slower because in the second and third layers there are filters of different sizes, causing some troubles for parallel computing. The decoding time using CNNs does not meet real-time requirement on current main-stream personal computers, which calls for further efforts on optimizing the computational architecture.

\begin{table}
\begin{center}
\caption{The results of decoding time (seconds per frame) of AR-CNN, VDSR and VRCNN} \label{Tab:time}
\begin{tabular}{|l|c|c|}
 \hline
  \multirow{2}{*}{Network} & \multicolumn{2}{c|}{Mode}\\
  \cline{2-3}
  &CPU&GPU\\
 \hline
 AR-CNN& 0.72 &0.33\\
 \hline
 VDSR&2.15 &1.27\\
 \hline
 VRCNN (ours)&0.98 &0.45\\
 \hline
\end{tabular}
\end{center}
\end{table}
Last but not the least, since the trained CNN is used for post-processing, especially at the decoder side, its memory cost is an important issue. We also compare the sizes of trained networks of AR-CNN, VDSR, and our VRCNN. The results are given in Table \ref{Tab:size}. Obviously, our VRCNN requires the lowest memory cost on storing the network because it is much shallower than VDSR and also has much less parameters than AR-CNN (shown in Table \ref{tab:para-arcnn} and Table \ref{tab:para}).
\begin{table}
\begin{center}
\caption{The sizes of trained networks (number of bytes required to store) of AR-CNN, VDSR and VRCNN} \label{Tab:size}
\begin{tabular}{|l|c|}
 \hline
  Network & Size\\
  \hline
 AR-CNN &  417 KB \\
 \hline
  VDSR & 2600 KB \\
 \hline
 VRCNN (ours)  &  214 KB \\
 \hline
\end{tabular}
\end{center}
\end{table}

\section{Conclusion}
\label{sec:conclusion}

In this paper, we have presented a convolutional neural network for post-processing in HEVC intra coding. The proposed network VRCNN outperforms the previously studied AR-CNN or VDSR in achieving higher bit-rate reduction, lower memory cost, and multiplied computational speedup. Compared to the HEVC baseline, VRCNN achieves on average 4.6\% BD-rate (in luminance) on the HEVC standard test sequences. Our future work is planned in two directions. First, we will extend VRCNN for HEVC inter coding, i.e. processing P and B frames. Second, we will investigate how to further simplify the network while maintaining its coding efficiency.

\section*{Acknowledgment}
This work was supported by the National Program on Key Basic Research Projects (973 Program) under Grant 2015CB351803, by the Natural Science Foundation of China (NSFC) under Grant 61331017, Grant 61390512, and Grant 61425026, and by the Fundamental Research Funds for the Central Universities under Grant WK2100060011 and Grant WK3490000001.


\begin{thebibliography}{10}
\providecommand{\url}[1]{\texttt{#1}}
\providecommand{\urlprefix}{URL }

\bibitem{bjontegaard2001calcuation}
Bjontegaard, G.: Calcuation of average {PSNR} differences between {RD}-curves.
  VCEG-M33 (2001)

\bibitem{dong2015compression}
Dong, C., Deng, Y., Loy, C.C., Tang, X.: Compression artifacts reduction by a
  deep convolutional network. In: ICCV. pp. 576--584 (2015)

\bibitem{dong2014learning}
Dong, C., Loy, C.C., He, K., Tang, X.: Learning a deep convolutional network
  for image super-resolution. In: ECCV. pp. 184--199. Springer (2014)

\bibitem{fu2012sample}
Fu, C.M., Alshina, E., Alshin, A., Huang, Y.W., Chen, C.Y., Tsai, C.Y., Hsu,
  C.W., Lei, S.M., Park, J.H., Han, W.J.: Sample adaptive offset in the {HEVC}
  standard. IEEE Transactions on Circuits and Systems for Video Technology
  22(12),  1755--1764 (2012)

\bibitem{girshick2014rich}
Girshick, R., Donahue, J., Darrell, T., Malik, J.: Rich feature hierarchies for
  accurate object detection and semantic segmentation. In: CVPR. pp. 580--587
  (2014)

\bibitem{he2016deep}
He, K., Zhang, X., Ren, S., Sun, J.: Deep residual learning for image
  recognition. In: CVPR. pp. 770--778 (2016)

\bibitem{jia2014caffe}
Jia, Y., Shelhamer, E., Donahue, J., Karayev, S., Long, J., Girshick, R.,
  Guadarrama, S., Darrell, T.: Caffe: Convolutional architecture for fast
  feature embedding. In: ACM Multimedia. pp. 675--678. ACM (2014)

\bibitem{kim2016accurate}
Kim, J., Lee, J.K., Lee, K.M.: Accurate image super-resolution using very deep
  convolutional networks. In: CVPR. pp. 1646--1654 (2016)

\bibitem{krizhevsky2012imagenet}
Krizhevsky, A., Sutskever, I., Hinton, G.E.: Imagenet classification with deep
  convolutional neural networks. In: NIPS. pp. 1097--1105 (2012)

\bibitem{nair2010rectified}
Nair, V., Hinton, G.E.: Rectified linear units improve restricted {Boltzmann}
  machines. In: International Conference on Machine Learning (ICML). pp.
  807--814 (2010)

\bibitem{norkin2012hevc}
Norkin, A., Bjontegaard, G., Fuldseth, A., Narroschke, M., Ikeda, M.,
  Andersson, K., Zhou, M., Van~der Auwera, G.: {HEVC} deblocking filter. IEEE
  Transactions on Circuits and Systems for Video Technology  22(12),
  1746--1754 (2012)

\bibitem{park2016cnn}
Park, W.S., Kim, M.: {CNN}-based in-loop filtering for coding efficiency
  improvement. In: 2016 IEEE 12th Image, Video, and Multidimensional Signal
  Processing Workshop (IVMSP). pp. 1--5. IEEE (2016)

\bibitem{sullivan2012overview}
Sullivan, G.J., Ohm, J.R., Han, W.J., Wiegand, T.: Overview of the high
  efficiency video coding ({HEVC}) standard. IEEE Transactions on Circuits and
  Systems for Video Technology  22(12),  1649--1668 (2012)

\bibitem{sun2003image}
Sun, J., Zheng, N.N., Tao, H., Shum, H.Y.: Image hallucination with primal
  sketch priors. In: CVPR. vol.~2, pp. 729--736. IEEE (2003)

\bibitem{szegedy2015going}
Szegedy, C., Liu, W., Jia, Y., Sermanet, P., Reed, S., Anguelov, D., Erhan, D.,
  Vanhoucke, V., Rabinovich, A.: Going deeper with convolutions. In: CVPR. pp.
  1--9 (2015)

\bibitem{wallace1992jpeg}
Wallace, G.K.: The {JPEG} still picture compression standard. IEEE Transactions
  on Consumer Electronics  38(1),  xviii--xxxiv (1992)

\bibitem{wang2016d3}
Wang, Z., Chang, S., Liu, D., Ling, Q., Huang, T.S.: D3: Deep dual-domain based
  fast restoration of {JPEG}-compressed images. In: CVPR. pp. 2764--2772 (2016)

\bibitem{xie2015holistically}
Xie, S., Tu, Z.: Holistically-nested edge detection. In: ICCV. pp. 1395--1403
  (2015)

\end{thebibliography}
\end{document}